\documentclass[conference]{IEEEtran}
\IEEEoverridecommandlockouts
\usepackage{cite}
\usepackage{amsmath, amssymb, amsfonts}
\usepackage{graphicx}
\usepackage{listings}
\usepackage{multirow}
\usepackage{url}
\usepackage{flushend}
\usepackage{subfig}
\usepackage{caption}

\makeatletter
\def\ps@IEEEtitlepagestyle{
  \def\@oddfoot{\mycopyrightnotice}
  \def\@evenfoot{}
}
\def\mycopyrightnotice{
  {\footnotesize
  \begin{minipage}{\textwidth}
  \centering
  \textcopyright 2022 IEEE.  Personal use of this material is permitted. Permission from IEEE must be obtained for all other uses, in any current or future media, including reprinting/republishing this material for advertising or promotional purposes, creating new collective works, for resale or redistribution to servers or lists, or reuse of any copyrighted component of this work in other works.
\end{minipage}
  }
}

\def\BibTeX{{\rm B\kern-.05em{\sc i\kern-.025em b}\kern-.08em
    T\kern-.1667em\lower.7ex\hbox{E}\kern-.125emX}}

\begin{document}

\title{Aspect-Oriented Programming based\\
building block platform to construct Domain-Specific Language for HPC application}

\author{\IEEEauthorblockN{1\textsuperscript{st} Osamu Ishimura}
\IEEEauthorblockA{\textit{Graduate School of Information Science and Technology} \\
\textit{The University of Tokyo}\\
Tokyo, Japan \\
oishimura@is.s.u-tokyo.ac.jp}
\and
\IEEEauthorblockN{2\textsuperscript{nd} Yoshihide Yoshimoto}
\IEEEauthorblockA{\textit{Graduate School of Information Science and Technology} \\
\textit{The University of Tokyo}\\
Tokyo, Japan \\
yosimoto@is.s.u-tokyo.ac.jp}
}

\maketitle

\begin{abstract}
The world of HPC systems is changing to a more complicated system because the performance improvement of processors has been slowed down. One of the promising approaches is Domain-Specific Language~(DSL), which provides a productive environment to create a high-efficient program without pain.
However, existing DSL platforms themselves often lack portability and cost DSL developers great effort.
To solve this issue, we propose an Aspect-Oriented Programming~(AOP) based DSL constructing platform, enabling developers to build a DSL platform by combining Aspect modules.
Aspect modules manage abstracted application flow, data structure, and memory access on our platform. 
Therefore, developers can create any DSL platform whose target application has the attributes which HPC applications usually have, the abstraction assumes.
This study implemented a prototype platform that can handle MPI and OpenMP layers. The prototype supports three types of applications (Structured-Grid, Unstructured-Grid, and Particle Simulation). 
Then, we evaluated the overheads caused by achieving flexibility and productivity of the platform.
\end{abstract}


\section{Introduction}

The recent tendency of the increasing complexity of structures of HPC systems causes the development of scientific applications that efficiently use HPC systems has become more painstaking.
It is because application developers must create both logic of the target application and logic to use an HPC system.

In a typical parallel-computed HPC application, the program is responsible for runtime initialization, data allocation, inter-process communication, and synchronization, whose complexities are proportional to a system's complexity.
Therefore, the more complex HPC systems become, the more complex the logic of programs will be.

However, this kind of logic, which causes the complication of programs, is not the core concern of the target application but the logic to efficiently use the HPC system.
Furthermore, these logics are scattered throughout the application, making it difficult to separate them as objects in Object-Oriented Programming.
In other words, they are ``cross-cutting concerns'' in Aspect-Oriented Programming~(AOP)~\cite{10.1007/BFb0053381} and are suitable for separated as \textit{Aspect}.

Many approaches have been simplified end-user programming with a Domain Specific Language~(DSL) in the HPC field.
DSLs can hide the cross-cutting concerns for using HPC systems from end-users.
However, these merely shift the responsibility to implement the cross-cutting concerns from end-users to DSL developers.
Furthermore, the situation is exacerbated by the fact that a DSL processing system needs to be more generic than a single application, making it more challenging to develop and maintain.

Most DSL processing systems for HPC, such as formura by T. Muranushi et al.~\cite{Muranushi2016b}, and Physis by N. Maruyama et al.~\cite{Maruyama2011} are designed only for specific combinations of computer system configurations, and porting to other platforms is not popular.
Since the lifetime of the program code is generally longer than the lifetime of the hardware, the lack of easy portability between hardware is a problem to be solved.
In previous studies, Legion~\cite{Bauer2014} and ebb~\cite{Bernstein2016} have made cross-cutting concerns separately or partially implementable.
However, they have not been developed, keeping the modularity and reusability of the optimization codes in mind.

How to divide the target problem at what granularity is important to realize the modularization with high reusability needs.
In most cases, a complex HPC system can be decomposed into several limited types of layers (such as distributed-memory parallel layers, shared-memory parallel layers, accelerator layers, and memory layers).
Therefore, if the necessary processing for each layer can be modelized into reusable AOP modules, and these modules can be used in any combination, portability among computer systems of DSL processing systems can be realized.

This research proposed a platform to create the DSL processing system for HPC applications. On the platform, developers can build DSL processing systems for specific HPC systems by combining AOP modules corresponding to the target HPC system hierarchy.
In addition, we implemented several DSL processing systems with different target applications on the prototype and evaluated the performance.

\section{Related Work}

As an attempt to remove the logic for HPC systems from the target application programs code using methods other than DSL, prior studies such as OpenMP~\cite{HomeOpen64:online}, OpenACC~\cite{Homepage32:online}, and XcalableMP~\cite{Xcalable30:online} use directives to append cross-cutting concerns to programs written in existing languages.
However, these directives-based approaches require modification of source codes.
Also, the logic inserted by directives is not reusable.

Existing methods for concealing differences at the intermediate language layer, like LLVM~\cite{TheLLVMC21:online} and Poly~\cite{PollyPol11:online}, the optimization infrastructure for LLVM, have a problem.
Translating program code into an intermediate language causes loss of meta information included in the logic of the target problem that could have been used for optimization.
A promising approach to solve this problem is MLIR~\cite{MLIR55:online}, which adds meta information to the intermediate language. 
If the DSL processing systems for HPC are implemented in MLIR, improved portability is likely to be realized.
However, it is still problematic because it requires DSL developers to learn the intermediate language, an entirely new language.

Other previous studies have applied AOP to scientific applications.
B. Harbulot and J. Gurd~\cite{Harbulot2006} developed LoopsAJ based on AspectJ~\cite{TheAspec64:online}, which is the AOP extension of Java.
LoopsAJ processes into Java bytecode and replaces the loop with parallel processing in Java Threads or MPJ (MPI-like Message Passing for Java)~\cite{MPJMPIli57:online}.
The work of J. Sobral~\cite{10.1007/978-3-540-71351-7_8} has also proposed an improved method to detect parallelizable loops from aspect runtime based on B. Harbulot and J. Gurd's work.
J. Dean~\cite{10.1145/2554850.2554917} proposed a model for parallelizing sequential applications with AOP and evaluated it in an implementation on AspectJ.

In our study, we focus on the characteristics of HPC applications and the typical structures of HPC systems and propose a model for parallelization and a method for creating modules with high reusability with reference to the findings by previous studies of what \textit{Pointcuts} to make and where to make them. 

\section{Proposed Platform}

\subsection{Used Technology}

\subsubsection{Aspect-Oriented Programming~(AOP)}
\label{sec:aop}

Aspect-Oriented Programming is an extension of Object-Oriented Programming~(OOP), a programming paradigm that separates cross-cutting concerns that OOP cannot separate as \textit{Aspect}, proposed by G. Kiczales et al.~\cite{10.1007/BFb0053381}.
The general behavioral model of AOP is the JoinPoint Model~(JPM), which is a model for inserting (or overriding) additional processes called \textit{Advice} into \textit{JoinPoints} generated by pattern matching called \textit{Pointcut}.
There are several ways to insert \textit{Advice}: before, after, or replacing the entire process.
JPM has the advantage of reusing aspects that can be applied for other programs with the same pattern rather than separating cross-cutting concerns.

\subsubsection{AspectC++~(AC++)}

We used AspectC++, an AOP extension of C++ developed by O.Spinczyk et al.~\cite{10.5555/564092.564100}, to implement the prototype platform.
AC++ uses JPM. AC++ can generate \textit{JoinPoints} for both function calls (call) and function executions (execution).

\subsection{Design of Platform}

\subsubsection{Basic Assumption}

We assume that the following three assumptions are often met in scientific applications to make the platform efficient in the designing platform.

\begin{LaTeXdescription}
\item[\textrm{Assumption I}]\mbox{}\\
The logic of the application is iterative, and the data at step $n$ is obtained by performing some arithmetic operation (called \textit{kernel}) on the data at step $n-1$.
\item[\textrm{Assumption II}]\mbox{}\\
The memory access pattern of the \textit{kernel} determined by the input data is generally static at each iteration, and the number of times the memory access pattern is changed is a sufficiently small ratio compared to the number of iterations.
\item[\textrm{Assumption III}]\mbox{}\\
The computation in the application has spatial locality.
\end{LaTeXdescription}

HPC scientific applications in each category of the seven dwarfs~\cite{colella2004defining}  generally satisfy the above assumptions.

\subsubsection{Execution Model}

In general, HPC systems have a hierarchical structure.

In a typical configuration, multiple nodes are connected by a network, each node has multiple CPUs, and each CPU has multiple cores. It may also have accelerators.
This variety of configurations is built by combining a relatively small number of types at each layer.
In other words, it is possible to create reusable modules for each layer of the hierarchy and significantly reduce development costs by making it possible to use these modules in combination.

In order to achieve this, our platform uses a task-based parallel programming model.
Multiple tasks run parallel on the system, and the final result is obtained by coordinating the tasks assigned to them.
This model is similar to the case of parallelization with spatial blocking on a distributed-memory parallel system.
In the basic execution model of parallelization with spatial blocking on a distributed-memory parallel system, each process is allocated one program and one blocked data area according to the number of processes. Then each process repeatedly updates blocked data.
On the other hand, in our platform, the area to be computed is blocked into data areas of fixed size(called \textit{Block}), and each task is allocated multiple \textit{Blocks}.
Each task iteratively updates each \textit{Block} with a \textit{subkernel} that updates only one \textit{Block}, as opposed to a \textit{kernel} that updates with the entire data.

In multi-layer parallelization, the module corresponding to each layer splits the \textit{Blocks} allocated by the upper layer into multiple and reallocates them to the layers of the lower layer.

\begin{figure}
\includegraphics[width=\linewidth]{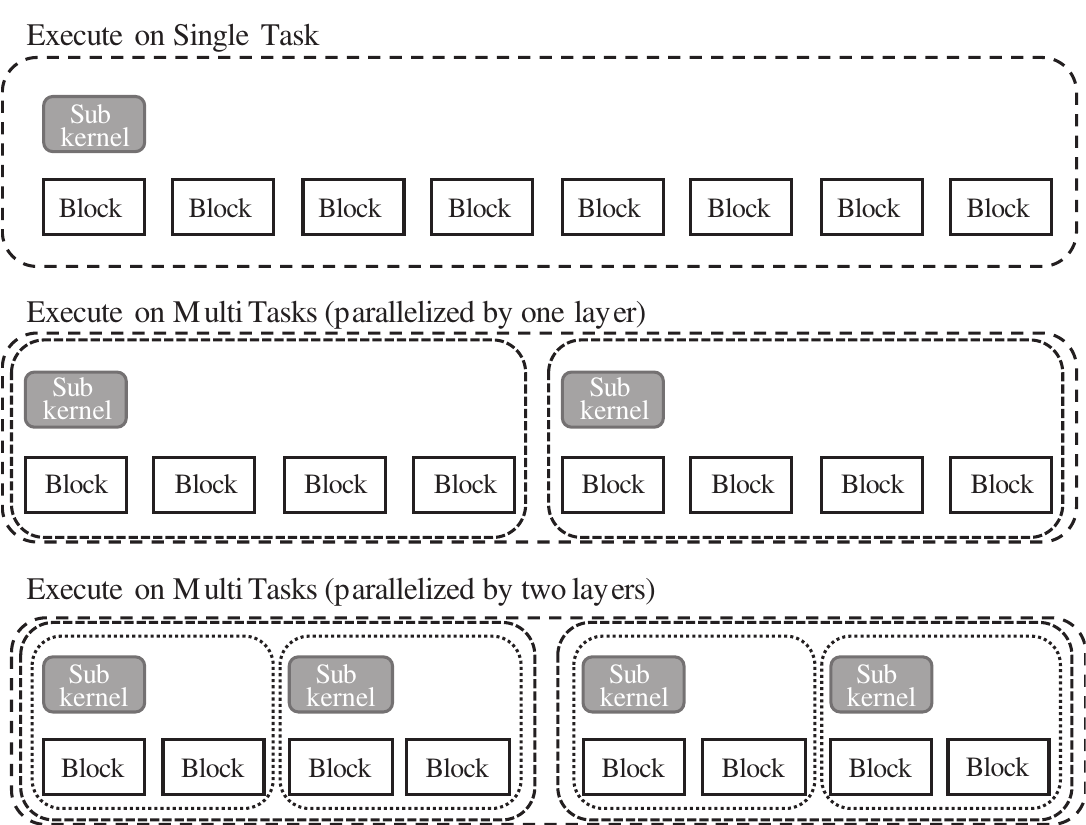}
\caption{Execution model of our platform. Each task surrounded by a dotted border is a task generated by each hierarchical recursive partitioning.}
\label{fig:model}
\end{figure}

Fig.~\ref{fig:model} shows the overview of this model.
End-users create a program that runs on a single task using the library provided by the DSL created by the platform.
On the program, the aspect module intervenes in controlling runtime initialization, \textit{Block} allocation, inter-process communication, and synchronization performed by the library to execute parallel operations.

\subsubsection{Data Model}

The global structure of the target data is represented by a tree structure of \textit{Blocks}~(\textit{Env}).
A \textit{Block}, which is a unit of data to be computed by a \textit{subkernel}, is a fixed-size data structure with dimensions implemented for each target computation.
For example, if a \textit{Block} is designed for a two-dimensional structured grid, the \textit{Block} is a small two-dimensional grid that divides a two-dimensional region.
Every \textit{Block} has its placement information in space.

A \textit{Block} may be an entity \textit{Block} with data~(\textit{Data Block}) or a virtual one~(\textit{Virtual Block}).
\textit{Data Block} has a multi-buffering to store the data. 
The buffers consist of a collection of memory chunks.
It is possible to combine memory chunks obtained from multiple pools.
The structure aims to handle non-uniform memory layers and memory-mapped files that appear in HPC applications with the same interface.

Additionally, \textit{Block} has parameters shown below
\begin{itemize}
\item  isValid: Indicates if the data is readable.
\item  data\_manage\_task\_id~(\textit{dm\_tid}): ID of the task in charge of initialization, update of multi-buffering, communication and finalization.
\item  calc\_handle\_task\_id~(\textit{ch\_tid}): ID of the task in charge of computation.
\end{itemize}

As \textit{Virtual Blocks} other than \textit{Data Blocks}, there are 
\textit{Empty Blocks} (which is just for joint of tree), 
\textit{Buffer-only data Blocks} (which is buffer for data communication from other tasks), 
\textit{Static Data Blocks} that provides static data, 
\textit{Arithmetic Blocks} that generate data by arithmetic expressions (used for Dirichlet boundary condition), 
and \textit{Reference Blocks} that reference other \textit{Blocks} (used for Neumann boundary condition).
Note that only \textit{Data Blocks} have valid \textit{dm\_tids} and are assigned to the task as the target of calculation.

The structure of this \textit{Env} is held in distribution among tasks.
In the case of a shared memory environment, tasks share the \textit{Env} save the memory usage.

\begin{figure}
  \begin{tabular}{cc}
    \begin{minipage}[t]{0.47\hsize}
      \centering
      \subfloat[]
      {
        \includegraphics[width=\linewidth]{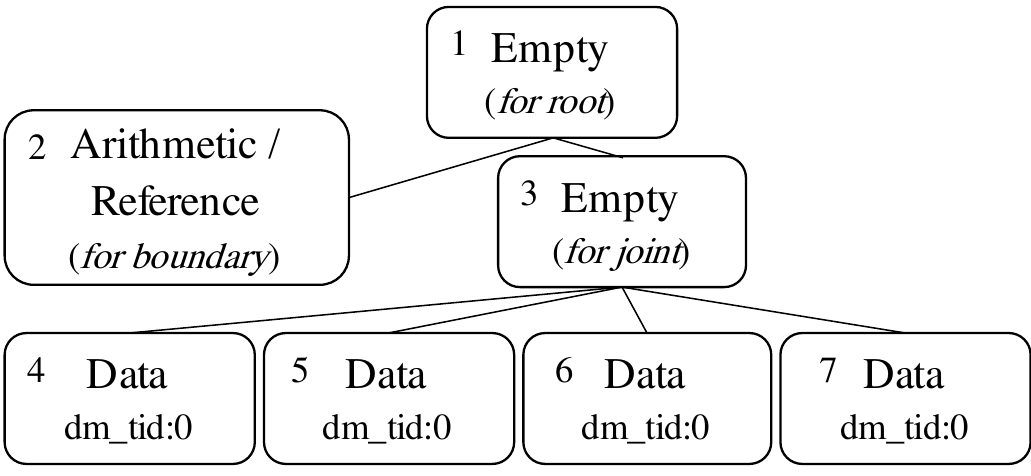}
        \label{fig:envtree_a}
      }
     \end{minipage} &
     \begin{minipage}[t]{0.47\hsize}
      \parbox{\linewidth}{
      \footnotesize
      (a) Abstract structure of Env. \\
      (b) and (c) are initial structures \\
      of two Envs shared by task 1,2 (b) \\
      and task 3,4 (c).}
    \end{minipage} \\[6ex]
    \begin{minipage}[t]{0.47\hsize}
      \centering
      \subfloat[]
      {
        \includegraphics[width=\linewidth]{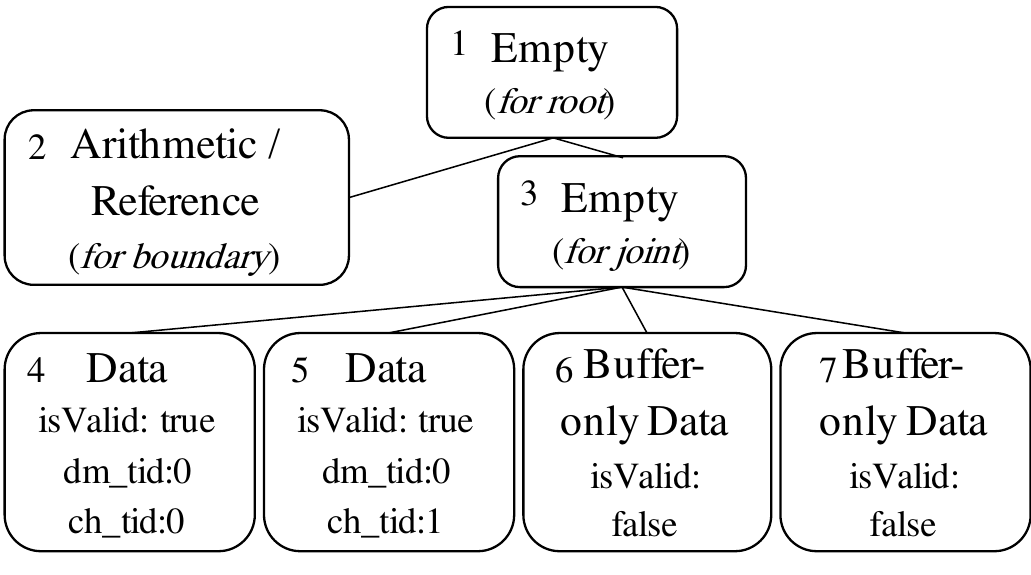}
        \label{fig:envtree_b}
      }
    \end{minipage} &
    \begin{minipage}[t]{0.47\hsize}
      \centering
      \subfloat[]
      {
        \includegraphics[width=\linewidth]{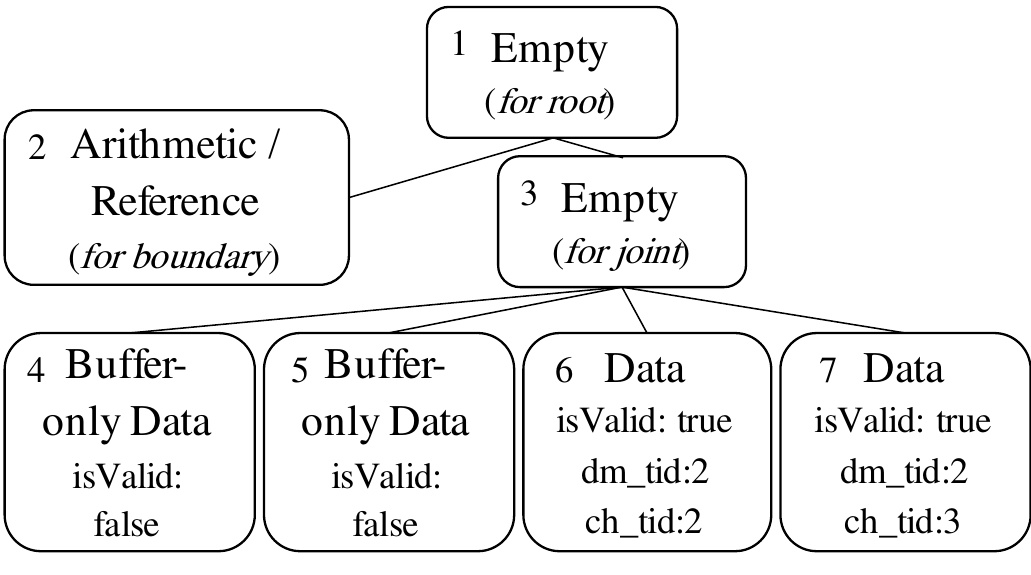}
        \label{fig:envtree_c}
      }
    \end{minipage}
  \end{tabular}
  \clearcaptionsetup{subfloat}
  \caption{\textit{Envs} on a pair of two shared memory tasks.}
\end{figure}

Fig.~\ref{fig:envtree_a} shows the example \textit{Env}, which run on a single task. 
By default, an \textit{Empty Blocks}; is a root of \textit{Env}, has a \textit{Block} for boundary, and \textit{Empty Blocks} as children. Then that \textit{Empty Blocks} has all \textit{Data Blocks} as children.
The reason for this structure is to improve the performance of searching for \textit{Blocks} with searched data.
In the case Assumption III, and if the tree is built to reproduce locality in space, the searching hits the target \textit{Blocks} faster by prioritizing sibling nodes and their children over its parent nodes.
Therefore, in the default tree, a \textit{Block} for the boundary is placed on different branches from the other \textit{Blocks} in terms of the root to hit it at last.
DSL developers can modify the tree by inserting \textit{Empty Blocks} below No.3 as new joints to increase locality to improve the performance of \textit{Env} search.
Fig.~\ref{fig:envtree_b} and Fig.~\ref{fig:envtree_c} are respective examples of Fig.~\ref{fig:envtree_a} when split on four tasks that share memory in a two-task set.
In the figure, only No.4 and No.5 are \textit{Data Blocks} assigned to tasks for calculation, and the others are \textit{Virtual Blocks}.
In the example, isValid flags of the \textit{Buffer-only data Blocks} are false, which become valid after the data is retrieved from other tasks on demand during the execution.

\subsubsection{Overview of the Platform}
In this study, we implement a platform for internal DSLs embedded within host languages and use syntaxes of host languages instead of external DSLs, which have their syntaxes and processing system, to increase the portability of existing codes and reduce the learning cost for end-users.
One disadvantage is that there is little gain to reduce the number of lines of code as a DSL.
The code related to the platform consists of the following three parts.

\begin{enumerate}
\item[A.] Platform Part: Libraries provided by the platform
  \begin{enumerate}
  \item[1.] Annotation Library (C++)
  \item[2.] Memory Library (C++)
  \item[3.] Aspect Module Library (AC++)
  \end{enumerate}
\item[B.] DSL Part: Libraries for the target application created by DSL developers
  \begin{enumerate}
  \item[1.] Annotation Library for Target Apps (C++)
  \item[2.] Memory Library for Target Apps (C++)
  \end{enumerate}
\item[C.] App Part: Codes created by end-users.
  \begin{enumerate}
  \item[1.] Application Code (C++)
  \end{enumerate}
\end{enumerate}



\begin{figure}
  \includegraphics[width=\linewidth]{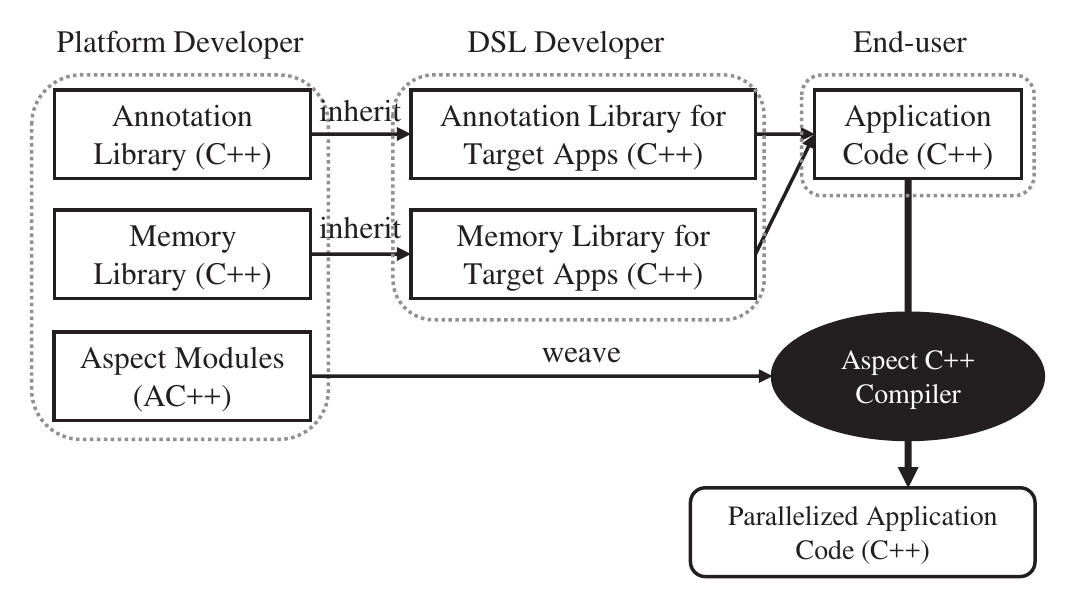}
  \caption{Overview of the elements of the platform and their respective developers.}
  \label{fig:compile}
\end{figure}

An end-user selects a class suitable for the target application from the library provided by the DSL, inherits the class, and creates the application code according to the platform's coding rule.
End-users compile their ``Application Code'' by C++ compiler into binary for serial operation.
For parallel operation, end-users select Aspect modules that work on the target system transcompile the Application Code and the Aspect Modules to a C++ source code for HPC systems using the AC++ compiler.
It is summarized as shown in Fig.~\ref{fig:compile}.

\subsubsection{Annotation Library}

Using the JPM model to isolate cross-cutting concerns has the advantage described in section~\ref{sec:aop}, but it also means that if a \textit{Pointcut} with a generic pattern is provided, there is a risk that unintentional \textit{JoinPoints} will be created.
In order to avoid this issue, the platform defines \textit{Pointcuts} for the classes in the annotation library and memory library and inherits classes of them to avoid the above problem. 
The details of \textit{Pointcut} implementation are described in the section~\ref{sec:ams}.

In the virtual class provided by the annotation library, three functions are defined: \textit{Initialize}, \textit{Processing}, and \textit{Finalize}.
\textit{Initialize} is for initializing the data for the computation domain.
\textit{Processing} is for performing the steps of calculation.
\textit{Finalize} is for post-processing.

In turn, the platform executes these three functions in the class implemented by end-users by inheriting the virtual class.

\subsubsection{Memory Library}
\label{sec:mem}

The platform allocates a fixed-size memory(\textit{Memory Pool}), and the data for the computation domain is placed on it.
The memory library provides interfaces to access the data on it.

The Memory Library has a \textit{Block}-based interface and a \textit{Page}-based interface.

The \textit{Block}-based interface provides an interface for end-user programs with data access in Block units of the structure implemented in DSL Part.
The \textit{Page}-based interface provides an interface for aspect modules to manage memory state regarding validness and dirtiness.
By converting the data accesses in end-user programs specified in units defined in the DSL Part to per-\textit{Page} accesses, the Memory Library and DSL Part can reuse Aspect Modules.
Note that a single page can contain multiple units of data. (For example, data of multiple points in a structured grid).
Furthermore, it enables to reduce the communication cost by communicating on a page-by-page basis compared to one on a block-by-block basis.

The Block-Based interface of the Memory Library is described below.

Firstly, \textit{Env} has an interface (named \textit{get\_blocks}) to extract the \textit{Blocks}, whose calc\_handle\_task\_id is caller's task id.

Secondly, the provided data update function (\textit{refresh}) tries to update the buffer.
The update succeeds if no access to non-existent data has been executed and returns ``true''.
This function is synchronously executed when there are multiple tasks.

The memory access interface of each \textit{Block} is a set of functions that access the data of the entire \textit{Env} starting from the corresponding \textit{Block}.
Data can be accessed using either the \textit{Global Address}, which represents the entire data area, or the \textit{Local Address}, which means the relative coordinates from the origin of each \textit{Block}.
When each memory access interface is called, if the target address is included in the \textit{Block} of the start point of the search, it accesses the data. Otherwise, it searches for the \textit{Block} that includes the address and accesses the data.
If no \textit{Block} contains the target address, it records the corresponding page number as a non-existent page.

The provided memory access interface can accept a flag statically or dynamically to indicate whether or not the data to be accessed undoubtedly contained in the \textit{Block} of the start point of the search as an argument.
If the flag is true, the platform executes the memory access without searching the environment tree.
On the other hand, if the flag is false, the platform searches the \textit{Env} for the \textit{Block} with the corresponding data and obtains the data through the found \textit{Block}.
From Assumption III, most memory accesses are in \textit{Blocks}, and \textit{Env} searches can be omitted, and DSL implementations or end-user applications can use it to speed up the program.

\begin{figure}
  \begin{tabular}{c}
    \begin{minipage}[t]{0.9\hsize}
      \subfloat[][The flow of memory accesses.]{
        \includegraphics[width=0.95\linewidth]{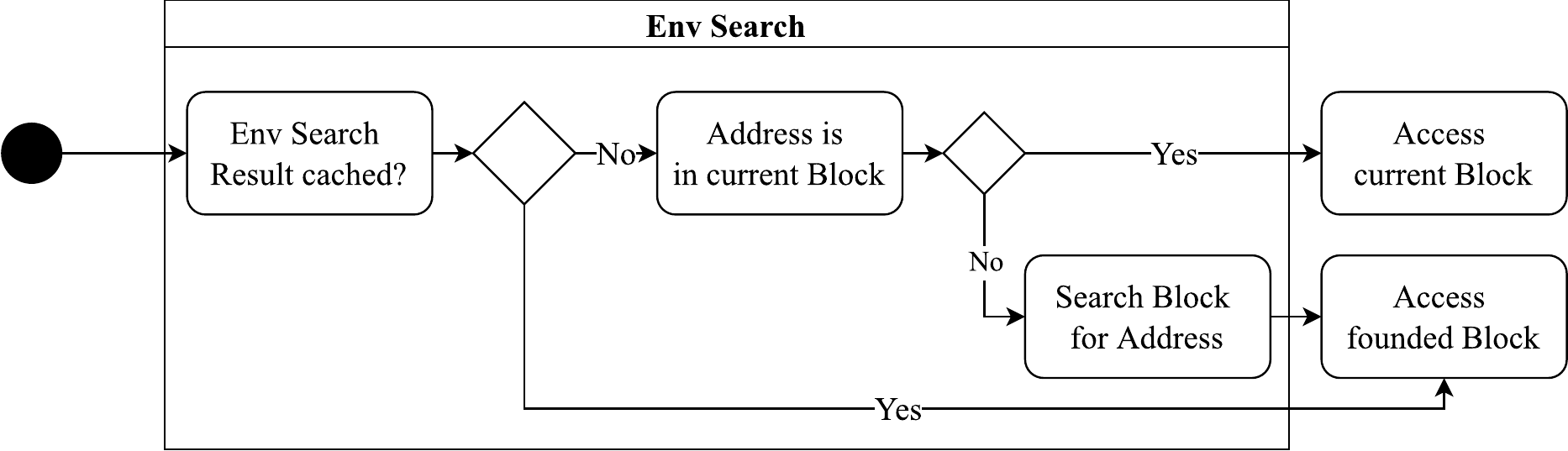}
        \label{fig:flow1}
      }
    \end{minipage}\\
    \begin{minipage}[t]{0.9\hsize}
      \subfloat[][The flow of memory accesses with MMAT.]{
        \includegraphics[width=0.95\linewidth]{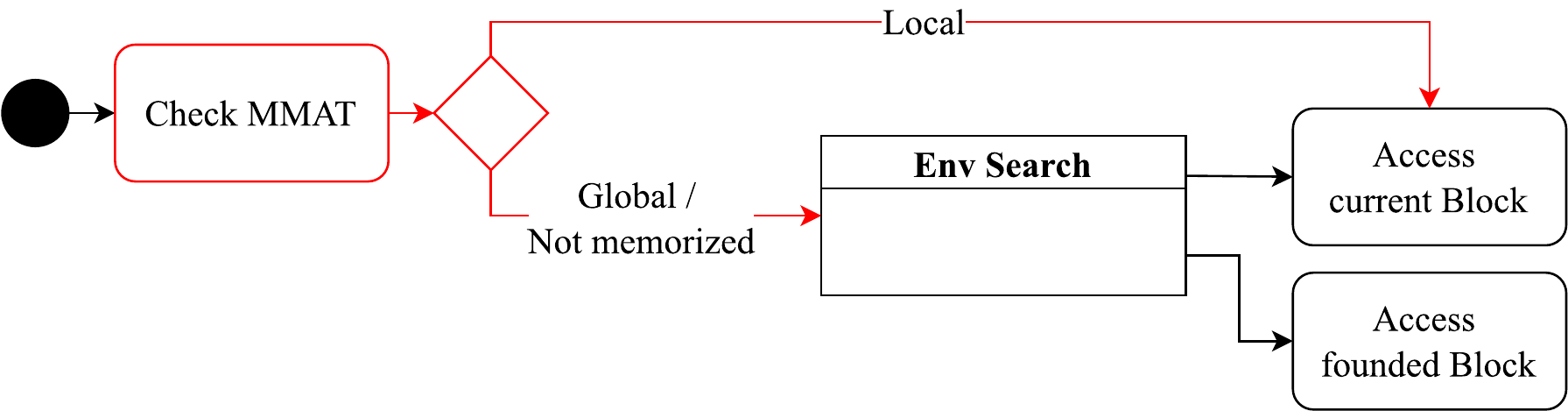}
        \label{fig:flow2}
      }
    \end{minipage}
  \end{tabular}
  \caption{Comparison between flows of memory access with and without MMAT.}
\end{figure}

In addition, the platform has a function called Memorization of memory access type (MMAT) that automates to omit \textit{Env} searches, and end-users can use this function by explicitly enabling it.
According to Assumption II, it is highly likely that the access pattern after the second step will be the same as the first step.
Therefore, by memorizing for each access, whether in- or out-of \textit{Block} access, it is possible to omit \textit{Env} search overheads.
Fig.~\ref{fig:flow1} and Fig.~\ref{fig:flow2} show how change the flow of memory access when MMAT is enabled.
This function does not automatically detect access pattern changes on the platform side, so end-users need to reset the MMAT when access patterns are changed.

\subsubsection{Aspect Module Structure}
\label{sec:ams}

The aspect module is a module that corresponds to each layer of an HPC system, and it manages the runtime of the corresponding layer.
In addition, the layer manages the number of parallelisms created in its layer, manages the number of tasks in the entire domain, and generates the task id of the corresponding layer based on the task id generated by the upper layer.
Furthermore, each master task of the corresponding layer synchronizes the data between the parallel tasks generated and processes data access requests from the upper layers.

Aspect modules are provided for each layer of the HPC system (MPI, OpenMP, etc.).
Each aspect module is composed of three main functions.

\begin{LaTeXdescription}
\item[\textrm{AspectType I}]\mbox{} Control of the runtime and tasks
\item[\textrm{AspectType II}]\mbox{} Assigning \textit{Blocks} to tasks
\item[\textrm{AspectType III}]\mbox{} Communication of data between tasks
\end{LaTeXdescription}

The logic is not implemented in some cases because it is unnecessary depending on the target runtime.
Each function is realized by sets of \textit{Pointcut} and \textit{Advice}.

First, for ``Control of the runtime and tasks'', the following are provided.

\begin{itemize}
\item \textit{Pointcut}: Entity point of program (\textit{main} function of C++ programs) and \textit{Initialize}, \textit{Processing}, and \textit{Finalize} of the virtual class of the Annotation Library.
\item \textit{Advice}: Initializing and Finalizing the target layer. Starting and terminating tasks. 
\end{itemize}

Second, for ``Assigning \textit{Blocks} to tasks'', the following are provided.

\begin{itemize}
\item \textit{Pointcut}: \textit{get\_blocks} of the Memory Library.
\item \textit{Advice}: Dividing the \textit{Blocks} allocated from the upper layer by the parallelism of the corresponding layer. 
\end{itemize}

Third, for ``Communication of data between tasks'', the following are provided.

\begin{itemize}
\item \textit{Pointcut}: \textit{refresh} of the Memory Library.
\item \textit{Advice}: Based on the list of non-existent page, get the data of the corresponding page from the task with the latest data. 
\end{itemize}

To improve communication performance, especially in a distributed-memory environment, a feature named ``Dry-run'' is implemented.
According to the platform specifications (Section~\ref{sec:app}), if the program access non-existent data, all results are discarded, and the \textit{Blocks} are retrieved from other tasks and recomputed.
However, in this case, in a distributed-memory environment, recalculation occurs at each step.
To solve this problem, we record the \textit{Blocks} that failed once and were acquired from other tasks and then acquire the data depending on the record before proceeding to the next step every time \textit{refresh} is called.
With Assumption 2, since the memory access pattern does not change significantly, a recalculation will be decreased by ``Dry-run''.

\subsubsection{Annotation Library for Target Apps and Memory Library for Target Apps}

Implements of DSL on the platform inherit the annotation and memory libraries provided by the platform, defines the types, and build the \textit{Env} necessary to implement the target application.

The structure of the \textit{Env} is defined in this part.
This part also defines the space coordinates and object IDs used in the application and defines how to map them to the \textit{Blocks} and tags for a \textit{Env} search.
Also, if the sugar syntax as a DSL is required, DSL developers implement it here.

\subsubsection{Application Design}
\label{sec:app}

As mentioned before, the DSL is provided as a C++ library on our platform.
The end-user inherits the class provided for the program they want to implement and writes logic using the types and variables defined in the superclass.
The implementation of the \textit{Processing} function must satisfy the following.

\begin{itemize}
\item After the completion of each step of the operation, the \textit{refresh} of the \textit{Env} defined in the class is called and proceeds to the next step only if the data update succeeds. If it fails, the step is re-executed.
\item In each step of the process, \textit{get\_blocks} function is called to get \textit{Blocks}, and each \textit{Block} is updated for one step in turn.
\end{itemize}

\section{Prototype Implementation}
\label{sec:pro}

\subsection{Aspect Module}

We have created aspect modules to manage OpenMP and MPI.
Then, the prototype platform can transform serial code to operate in three parallel runtime combinations: MPI, OpenMP, and MPI+OpenMP.
In the aspect of MPI, the initialization runtime and finalization runtime \textit{Advices} are performed before and after the entry point(\textit{main} of C++ programs) as AspectType I.

In the aspect of OpenMP, the starting tasks \textit{Advices} is performed before \textit{Processing} as AspectType I.
Moreover, AspectType III is not implemented because OpenMP is a shared-memory parallel system.

\subsection{Sample DSL Processing System}

We created a DSL system for three types of applications in 2D space: structured grid, unstructured grid, and particle method.
In the case of the particle method, only the case where the particles do not move between \textit{Blocks} is supported.

\subsection{Assigning \textit{Blocks} to Tasks}

On the prototype, the assignment of \textit{Blocks} to tasks is processed in the layer of the DSL processing system.
Each \textit{Data Block} inside the \textit{Env} has the index using Z-order curve~\cite{morton1966computer} (by using the PDEP instruction (x86)), and DSL processing systems assign \textit{Blocks} to each task depending on the index.

\subsection{Source Code Example}

An example of an application code that an end-user will write is shown in Listing~\ref{lst:usercode}.
Note that some variable names are shorted to fit this paper width.
(SIZE\_X to X, SIZE\_Y to Y, GlobalAddress\_t to GA\_t and LocalAddress\_t to LA\_t.)

\begin{lstlisting}[frame=single,caption=Example source code for 2D structured grid application on a DSL processing system.,captionpos=b,label=lst:usercode]
class Sample
 : SU_Target_SGrid2D<double, 8, 9> {
/*Variable declaration, Initialize 
and Finalize are omitted*/
void Processing() override {
 WarmUp(Kernel);
 for(int c=0; c<LOOPNUM; ++c) Run(Kernel);
}
template <bool isWarmUp> bool Kernel(){
 for(auto el:env.get_blocks<isWarmUp>()){
  InitKernelMacros(el);
  for(AddrE j=0; j<BLK::Y; ++j){
   for(AddrE i=0; i<BLK::X; ++i){
    double e_n, e_w, e, e_e, e_s;
    e_n=GetD(LA_t{{i, j-1}}, j > 0);
    e_w=GetD(LA_t{{i-1, j}}, i > 0);
    e=GetDD(LA_t{{i, j}});
    e_e=GetD(LA_t{{i+1, j}}, i+1<BLK::X);
    e_s=GetD(LA_t{{i, j+1}}, j+1<BLK::Y);
    auto ans=alpha*e 
        +beta*(e_e+e_w+e_s+e_n);
    SetD(LA_t{{i, j}}, ans);
   }
  }
 }
 return env.refresh<isWarmUp>();
}};
#define CUSTOM_TARGET Sample
\end{lstlisting}

Listing~\ref{lst:usercode} is an example of an App Part code.
In this application, end-users implemented Sample\_SGrid\_2d class inherited SU\_Target\_SGrid2D virtual class provided by DSL implementation.
The template parameter of SU\_Target\_SGrid2D is the number of adjacent points, the constant to determine the size of the \textit{Block} (it indicates $2^8\times2^8$), and the constant to determine the size of the page (it indicates $2^9$).
Three functions (\textit{Initialize}, \textit{Processing}, \textit{Finalize}) inherited from the parent class and a user-defined \textit{Kernel} are implemented as a end-users Application Code. 

In the \textit{Processing}, user should warm-up the \textit{Kernel} at first, then calculate the \textit{Kernel} $LOOPNUM$(a constant variable) times.

When calling the warm-up macro, \textit{Kernel} run multi-time with ``dry-run'' mode data access to gather the information of data communication among Blocks caused by the \textit{Kernel}. The previously collected information at ``MMAT'' is cleared when the warm-up macro is called.

\textit{Kernel} should receive a  bool variable as a template argument to switch between warm-up and execution.
Inside the \textit{Kernel} every \textit{Block} in the ``env'' are updated one by one. 

The \textit{Kernel} updates the data of the points.
To get / set the data, we use \textit{GetD}/\textit{GetDD} and \textit{SetD} with a relative address.
They are macros to simplify the expression, which are enabled by the \textit{InitKernelMacros} macro in advance.
\textit{GetDD} is an extension of \textit{GetD} and is given additional template parameter to skip data search mentioned at section \ref{sec:mem}.
After the update, the \textit{Kernel} call \textit{refresh} to switch write-buffers and read-buffers.
The return value of \textit{refresh} indicates the presence or absence of invalid data access. 
This value is passed to the platform as the return value of \textit{Kernel}, and the platform determines whether to forward the step or not.
\textit{CUSTOM\_TARGET} macro specifies the class to be executed by the platform.

\section{Evaluation}

The evaluation was conducted using the Fujitsu PRIMERGY CX400M1/CX2550M5 (Oakbridge-CX).

\subsection{Test Environment}

\begin{itemize}
\item CPU: Intel Xeon Platinum 8280 $\times$ 2
\item Memory (per node): 2,933 RDIMM 16 GB $\times$ 12
\item Network: Intel Omni Path (12.5 GB/s)
\item AC++ Compiler: AspectC++ compiler 2.3
\item C++ Compiler: Intel (ICC) 19.1.3.304 20200925
\end{itemize}

%

\subsection{Sample Program}
We created a simple benchmark software that models structured grid, unstructured grid, and particle method one without using the platform(``Handwritten'') and one with the platform(``Platform'').
The ``Handwritten'' code is a simple serial code with double-buffering without MPI, OpenMP, and SIMD optimization.

\begin{lstlisting}[frame=single,caption=Example source code of ``Handwritten'' for 2D structured grid application.,captionpos=b,label=lst:usercode_handwritten]
/*Initialize and Finalize are omitted*/
for(int c=0; c<LOOPNUM; c++){
  for(size_t y=0; y<mem.SIZE; ++y){
    for(size_t x=0; x<mem.SIZE; ++x){
    auto v1=alpha * mem.get(x, y);
    auto v2=beta * (mem.get(x-1, y)
                  + mem.get(x+1, y)
                  + mem.get(x, y-1)
                  + mem.get(x, y+1));
    mem.set(x, y, v1+v2);
    }
  }
  mem.refresh();
}
\end{lstlisting}
Listing~\ref{lst:usercode_handwritten} is the part of an example ``Handwritten'' code for a 2D structured grid application.
\textit{mem} is the double-buffering array implemented as a wrapper of std::vector array.
Inside the \textit{get} and \textit{set} function, address of the target data at std::vector array are calculated by the given two arguments.
Additionally, \textit{get} returns the value depending on the boundary condition if the address is outside the region.
\textit{refresh} exchanges the buffer.

\begin{figure}
  \begin{tabular}{cc}
    \begin{minipage}[t]{0.45\hsize}
      \centering
      \subfloat[][Example of a data layout and a grid data structure for the structured grid.]{
        \includegraphics[width=0.95\linewidth]{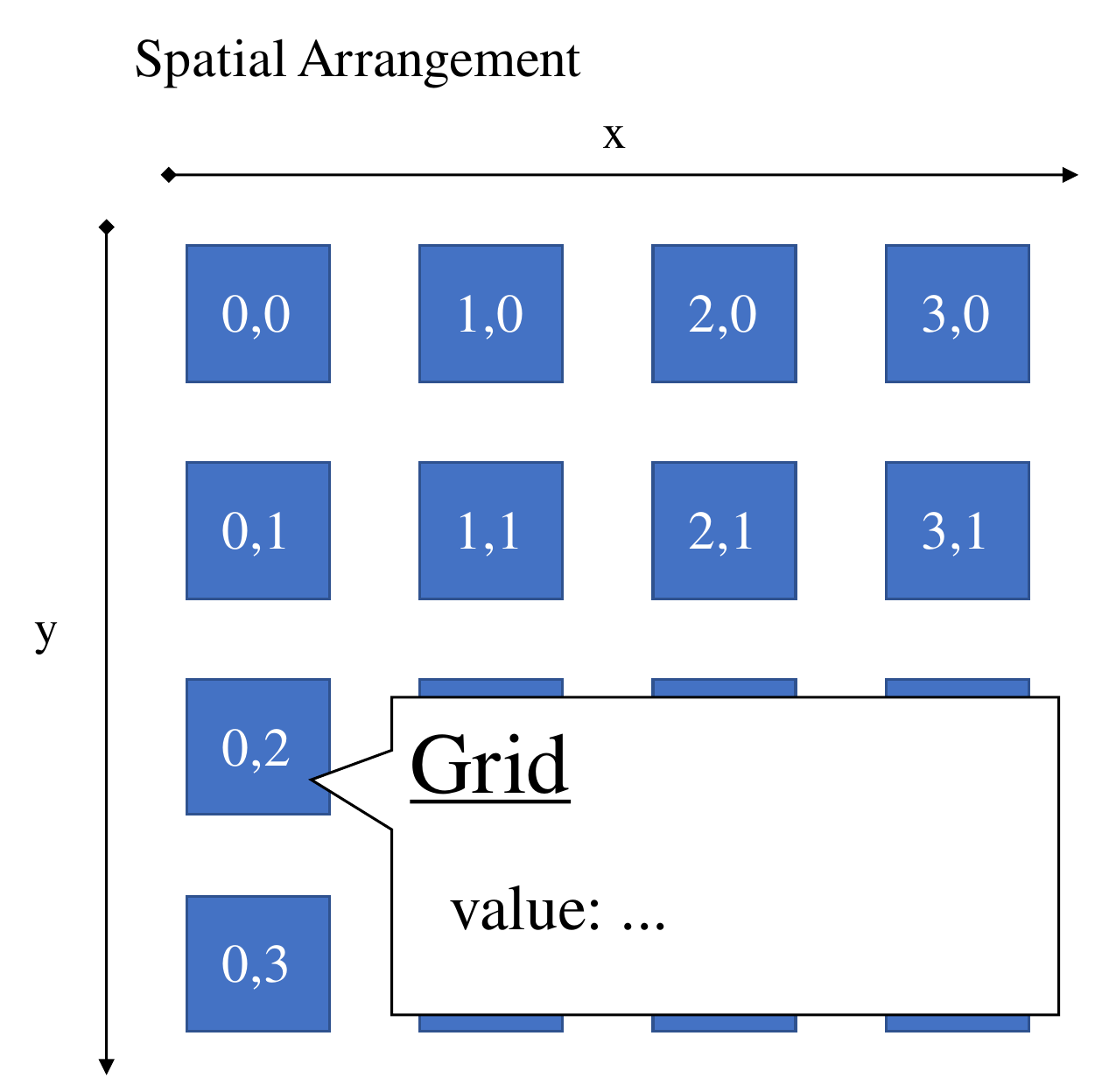}
        \label{fig:sgrid}
      }
    \end{minipage} &
    \begin{minipage}[t]{0.45\hsize}
      \centering
      \subfloat[][Example of a data layout and a grid data structure for the unstructured grid (CaseC).]{
        \includegraphics[width=0.95\linewidth]{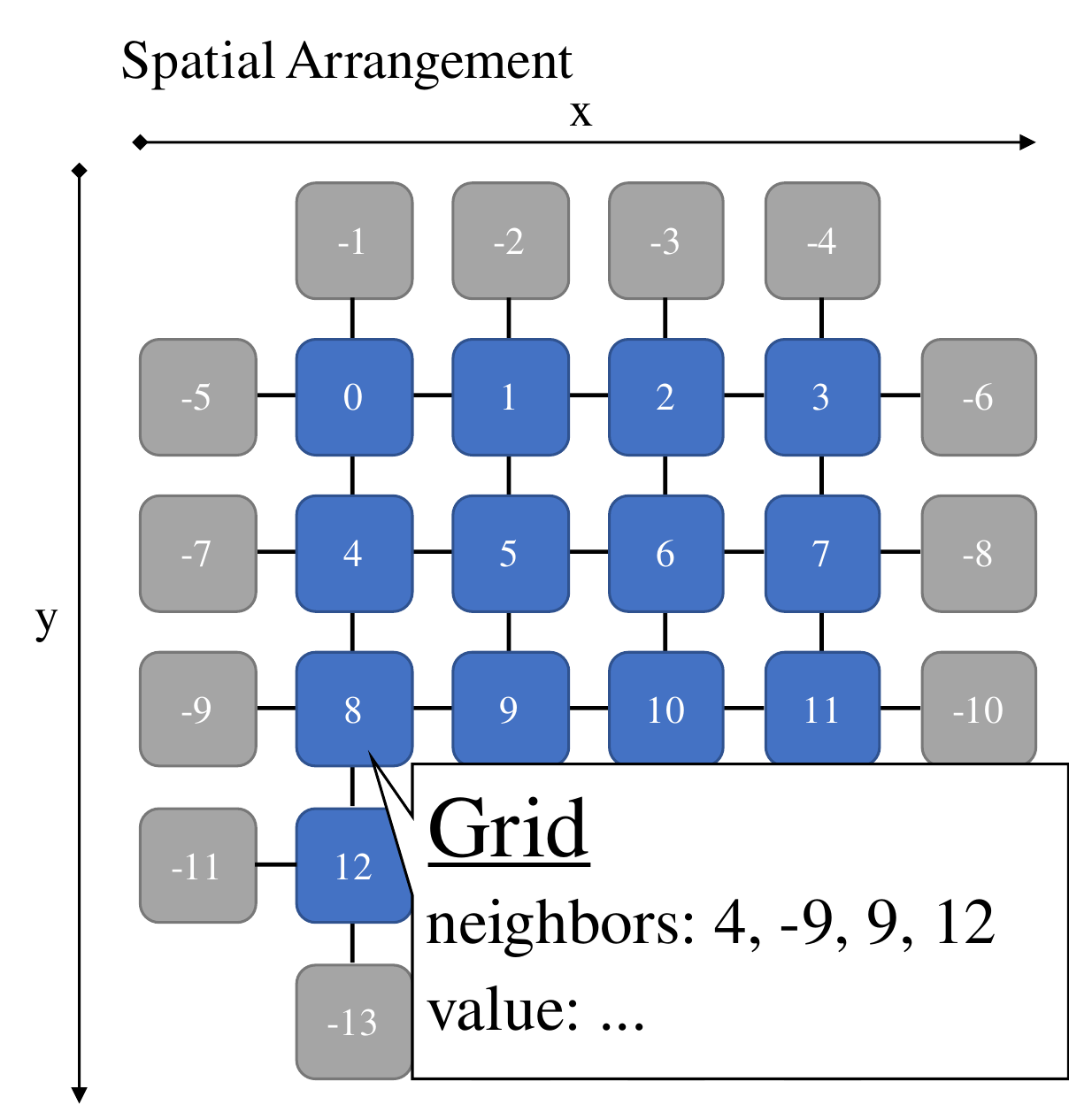}
        \label{fig:usgridc}
      }
    \end{minipage} \\ 
    \begin{minipage}[t]{0.45\hsize}
      \centering
      \subfloat[][Example of a data layout and a grid data structure for the unstructured grid (CaseR).]{
        \includegraphics[width=0.95\linewidth]{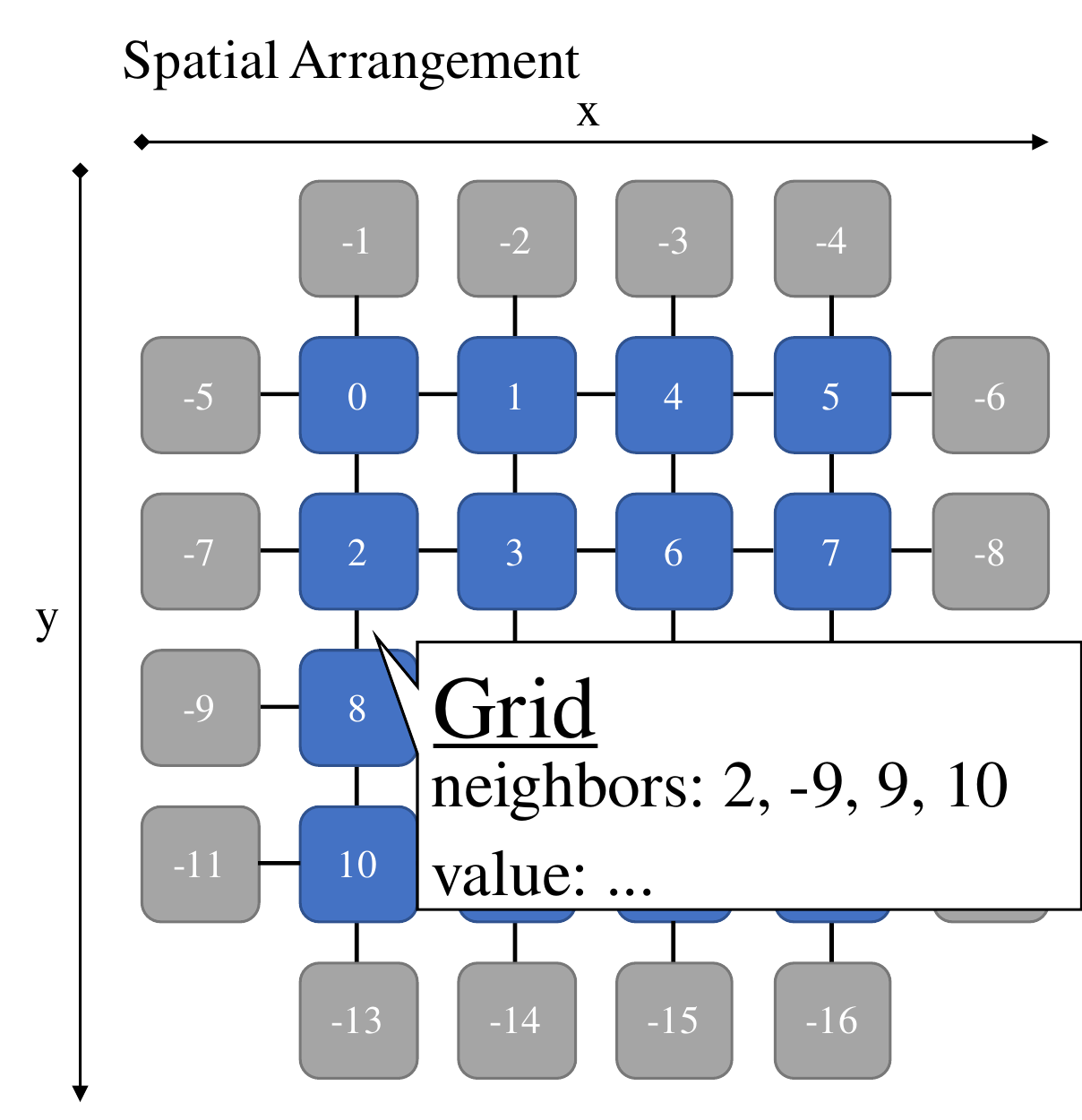}
        \label{fig:usgridr}
      }
    \end{minipage} & 
    \begin{minipage}[t]{0.45\hsize}
      \centering
      \subfloat[][Example of a data layout and a grid data structure for the particle simulation.]{
        \includegraphics[width=0.95\linewidth]{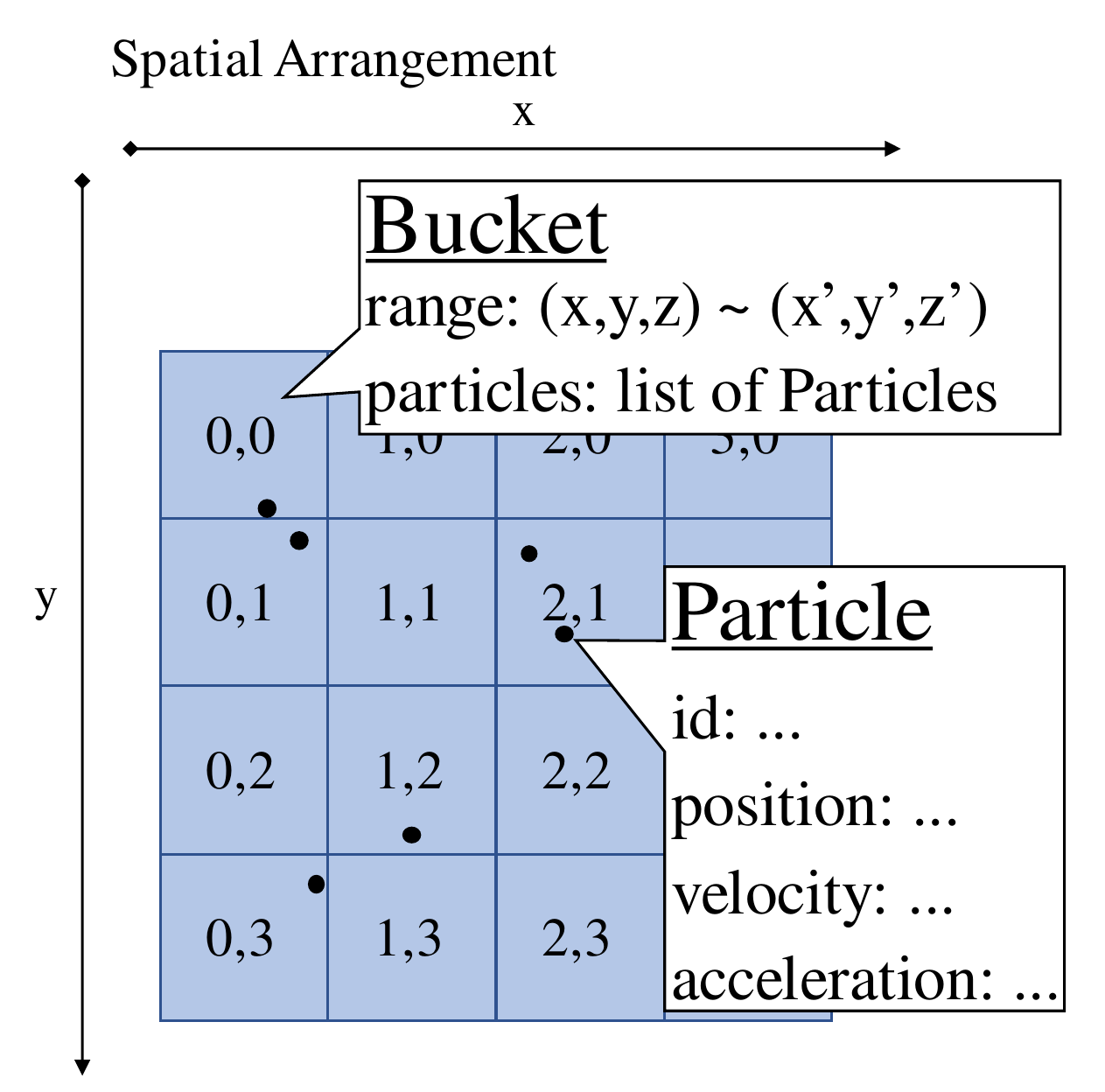}
        \label{fig:particle}
      }
    \end{minipage}
  \end{tabular}
  \caption{Data structure implementation of sample programs}
\end{figure}

\subsubsection{Sample Program for Structured Grid ``SGrid''}
This sample performs a simulation of a structured grid in a two-dimensional space.
The above space was regarded as a matrix for benchmarking purposes, and the Laplace equation with a 5-point finite difference scheme was solved by the Jacobi method.
Fig.~\ref{fig:sgrid} shows the example of a data layout and a grid data structure for the structured grid.
Each box represents the data on a grid point. Each vector2 value in the box indicates the address of the data.
Each data consists of the scalar value on the grid point.
Dirichlet boundary condition is set using \textit{Arithmetic Block} for the region outside the computational domain.

In the DSL Part of ``SGrid'', the parameters for the DSL processing system are defined as follows:
\begin{itemize} 
  \item \textit{Block} Size: $256\times256$
  \item \textit{Page} Size: $2^8$ pieces of data (16KB)
\end{itemize}

\subsubsection{Sample Program for Unstructured Grid ``USGrid''}
This sample performs a simulation of an unstructured grid in a two-dimensional space.
Unlike SGrid, USGrid has the Global Addresses of neighbors on the \textit{Env} as each data.
The above space was regarded as a matrix for benchmarking purposes, and the Laplace equation with a 5-point finite difference scheme was solved by the Jacobi method.
We prepared the following two cases with different memory accesses patterns for adjacent points.

\begin{itemize}
\item CaseC: Consecutive memory accesses with spatial locality (same as structured grid sample except for memory access to adjacent points are indirection references).
\item CaseR: Uncontiguous memory accesses. Do not have spatial locality. It does not satisfy Assumption III.
\end{itemize}
Fig.~\ref{fig:usgridc} and Fig.~\ref{fig:usgridr} show the example of a data layout and a grid data structure for the unstructured grid.
Each box represents the data on a grid point. Each number in the box indicates the address of the data.
Each data consists of the scalar value on the grid point and the list of the addresses of the neighbor grid points.
As a note, CaseC and CaseR have the same calculation, differing only in memory access.

Data outside the computational domain are placed at \textit{Static Data Block}.

In the DSL Part of ``USGrid'', the parameters for the DSL processing system are defined as follows:
\begin{itemize} 
  \item \textit{Block} Size: $256\times256$
  \item \textit{Page} Size: $2^8$ pieces of data (64KB)
\end{itemize}

\subsubsection{Sample program for particle method ``Particle''}
This sample performs the particle method in three-dimensional space but only one layer of particles in the z-axis.
Fig.~\ref{fig:particle} shows the example of a data layout and a particle data structure for particle simulation.
Each data represents the data of one bucket, and each bucket includes multiple particle data.

Each box represents the data on a bucket. Each number in the box indicates the address of the data.
Each dot represents the data on a particle.
Each data of a bucket consists of a list of the particles inside the bucket.
Each data of a particle consists of a particle id and three vector3 values(for a position, a velocity, and an acceleration).

The sample programs simulate the particles' behavior in a two dimensions space by setting the z-axis of space to one. 
It performs a particle simulation in a space whose boundaries are modeled by fixed wall particles.
The movable particles are placed in the interior space uniformly when it initialized.
Note that the current implementation does not implement the movement of particles between buckets. Therefore, the calculations are performed in the small case of time step and number of loops to avoid significant movement of particles.
When referring outside the computational domain, a block filled with dummy particles is returned by \textit{Arithmetic Block}.

In the DSL Part of ``Particle'', the parameters for the DSL processing system are defined as follows:
\begin{itemize}
  \item Number of buckets per \textit{Blocks}: $8\times8\times1$
  \item Number of particles per bucket: $16$
  \item \textit{Page} Size: $2^3$ pieces of data (about 12KB)
\end{itemize}

From the weight function of the influence distance between particles, the App Part can calculate the force by interacting with the particles in the surrounding eight buckets outside the target bucket.

\subsection{Overheads of platform}
\label{sec:overhead}

Firstly, we executed the three benchmarks with a single task and evaluated the overhead caused by the platform itself.
\begin{itemize}
\item Handwritten
\item Platform: Directly compiled as a simple C++ application by C++ compiler.
\item Platform NOP: Transcompiled through the AC++ compiler without aspects module.
\item Platform MPI: Transcompiled through the AC++ compiler with the MPI aspect module.
\item Platform OMP: Transcompiled through the AC++ compiler with the OpenMP aspect module.
\end{itemize}

For SGrid and Particle, it is possible to determine whether the access is inside or outside the \textit{Block} arithmetically using spatial information.
On the other hand, for USGrid, it is not possible.
Then we evaluate the performance without MMAT(``w/o MMAT'') and with MMAT(``w MMAT'') in these two cases.

The given parameters are shown below.
\begin{itemize}
\item RegionSize (``SGrid'', ``USGrid''): $2048\times2048$, $4096\times4096$ 
\item Number of Particles (``Particle''): $2^{16}$, $2^{18}$
\item Number of MPI Processes and OpenMP threads: $1$
\end{itemize}

The results are as shown in the figure. The graph shows the ratio when the performance of Handwritten is set to 100\%.
\begin{figure}
\includegraphics[width=0.95\linewidth]{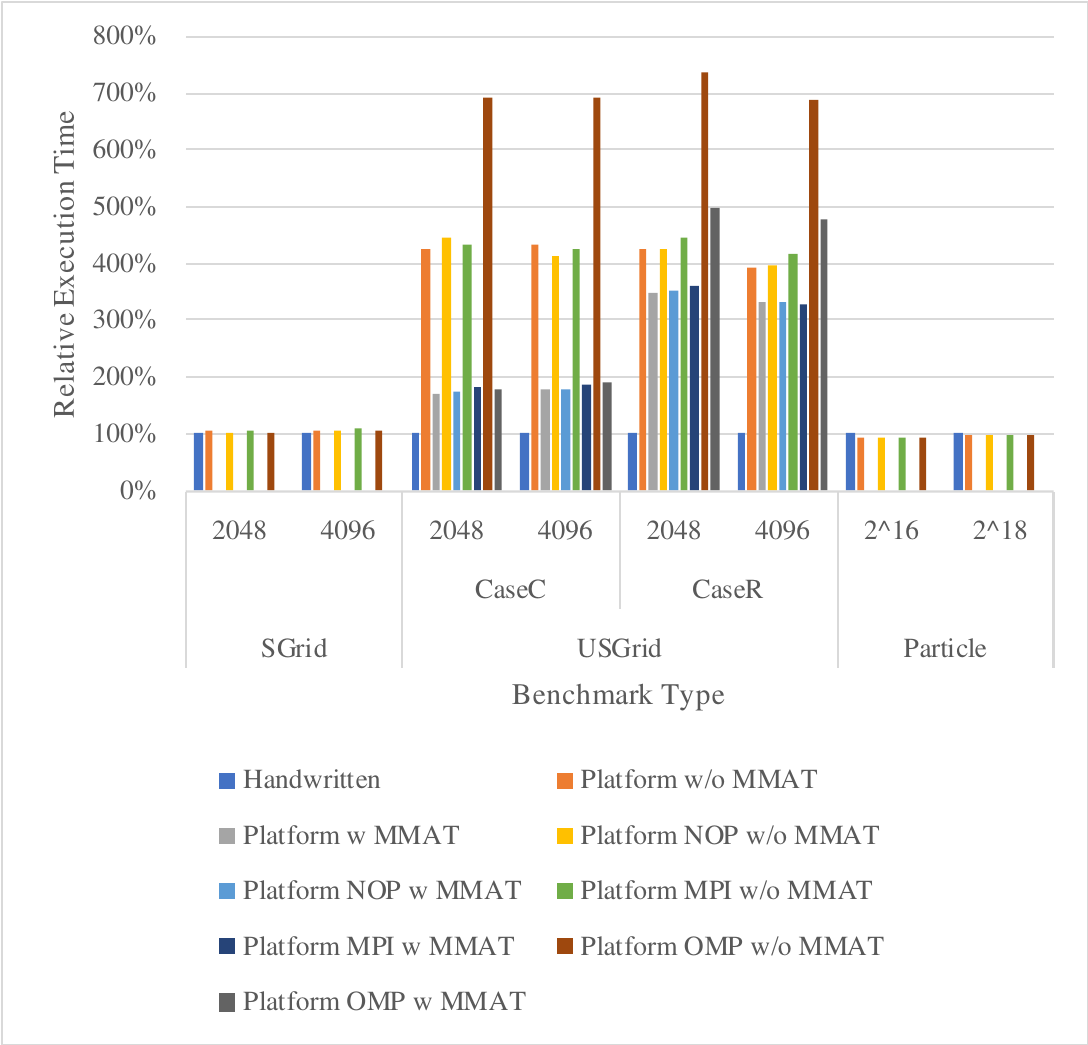}
\caption{Comparison of benchmark performance between application structures.}
\label{fig:bench1}
\end{figure}

Fig.~\ref{fig:bench1} indicates the following.

Firstly, the overhead due to the platform is maximally 600\%. However, the overheads can be reduced to minimally about 70\% using MMAT, depending on the access pattern.

Secondly, from the comparison between Platform and Platform NOP, the overhead due to the transcompilation through AspectC++ is about several percent.

Thirdly, the result of ``Particle'' indicates a slight performance improvement by porting to the platform caused by improved memory locality.

%

\subsection{Scaling}

In the benchmarks in this section, we used MMAT for USGrid samples.
In the graphs of Strong Scaling shown in this section, the execution times are normalized so that the time by one task becomes unity.
In the graphs of Weak Scaling shown in this section, the execution times are normalized so that the time by one task becomes 100\%.

\subsubsection{Strong Scaling (MPI)}

\begin{itemize}
  \item RegionSize (``SGrid'', ``USGrid''): $4096\times4096$ 
  \item Number of Particles (``Particle''): $2^{18}$
  \item MPI Process: One process per node
  \item Number of Process: 1, 2, 4, 8, and 16
\end{itemize}

\begin{figure}
\includegraphics[width=0.95\linewidth]{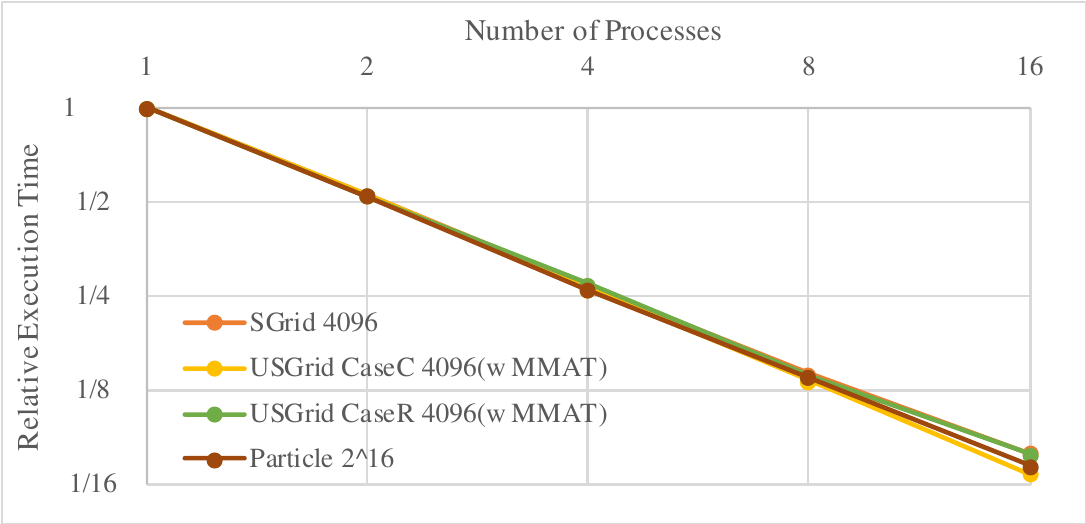}
\caption{Strong scaling performance on MPI parallel.}
\label{fig:bench2}
\end{figure}

Fig.~\ref{fig:bench2} indicates that the benchmark scaled almost linearly.

\subsubsection{Weak Scaling (MPI)}
\begin{itemize}
    \item Data Size (``SGrid'', ``USGrid''): $2048\times2048$ per task 
    \item Number of Particles (``Particle''): $2^{16}$ per task 
    \item MPI Process: One process per node
    \item Number of Process: 1, 4, 16, and 64 
\end{itemize}
\begin{figure}
\includegraphics[width=0.95\linewidth]{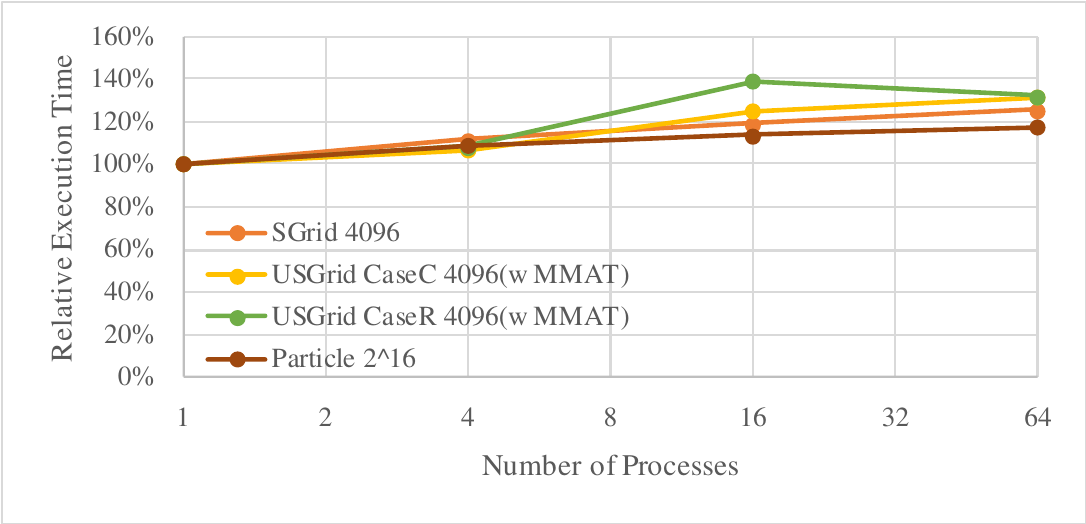}
\caption{Weak scaling performance on MPI parallel.}
\label{fig:bench3}
\end{figure}

Fig.~\ref{fig:bench3} indicates that the performance in ``USGrid CaseR'' is somewhat poor compared to the other cases. It is probably due to the significant communication overhead in ``USGrid CaseR'' since there are many accesses across \textit{Blocks} and the destination \textit{Blocks} are different.
However, the memory access pattern of ``CaseR'', shown in Fig.~\ref{fig:usgridr}, is not changed if the computational area is bigger enough.
Therefore, we have determined that the reason why the results of ``USGrid CaseR'' on 16 and 64 processes are almost the same might be the communication overhead for data exchange, which is the significant factor of performance degradation from 1 process to 16 processes, is the same for 16 and 64 processes.

\subsubsection{Strong Scaling (OpenMP)}
\begin{itemize}
  \item RegionSize (``SGrid'', ``USGrid''): $4096\times4096$
  \item Number of Particles (``Particle''): $2^{18}$
  \item Number of Thread: 1, 2, 4, 8, and 16 
\end{itemize}

\begin{figure}
\includegraphics[width=0.95\linewidth]{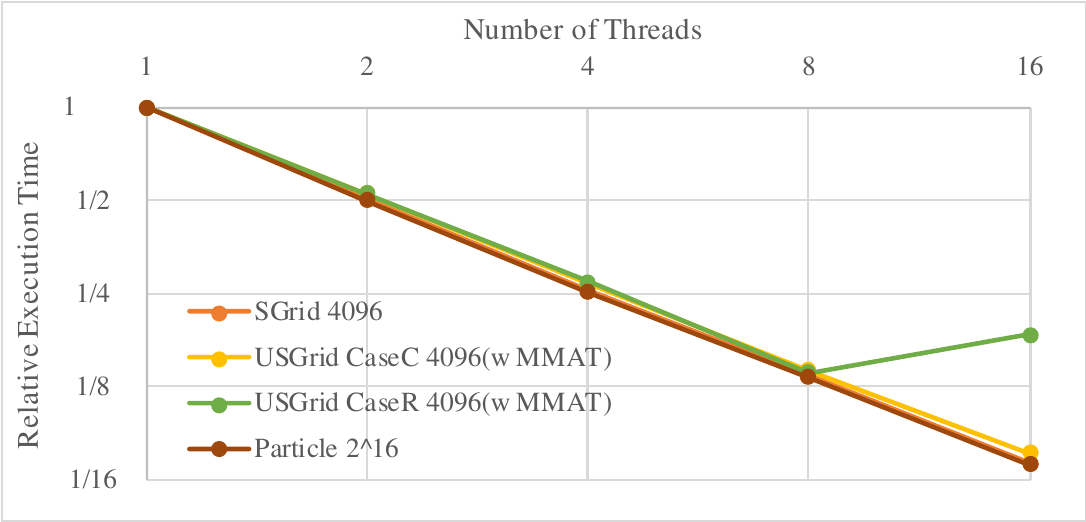}
\caption{Strong scaling performance on OpenMP parallel.}
\label{fig:bench4}
\end{figure}

Fig.~\ref{fig:bench4} indicates that except ``USGrid CaseR'' with 16 threads, The benchmark scaled almost linearly.
We guess that the result of ``USGrid CaseR'' is caused by less capacity of CPU cache and memory bandwidth per task.

\subsubsection{Weak Scaling (OpenMP)}
\begin{itemize}
  \item Data Size (``SGrid'' \& ``USGrid''): $2048\times2048$ per task 
  \item Number of Particles(``Particle''): $2^{16}$ per task 
  \item Number of Thread: 1, 4, and 16
\end{itemize}

\begin{figure}
\includegraphics[width=0.95\linewidth]{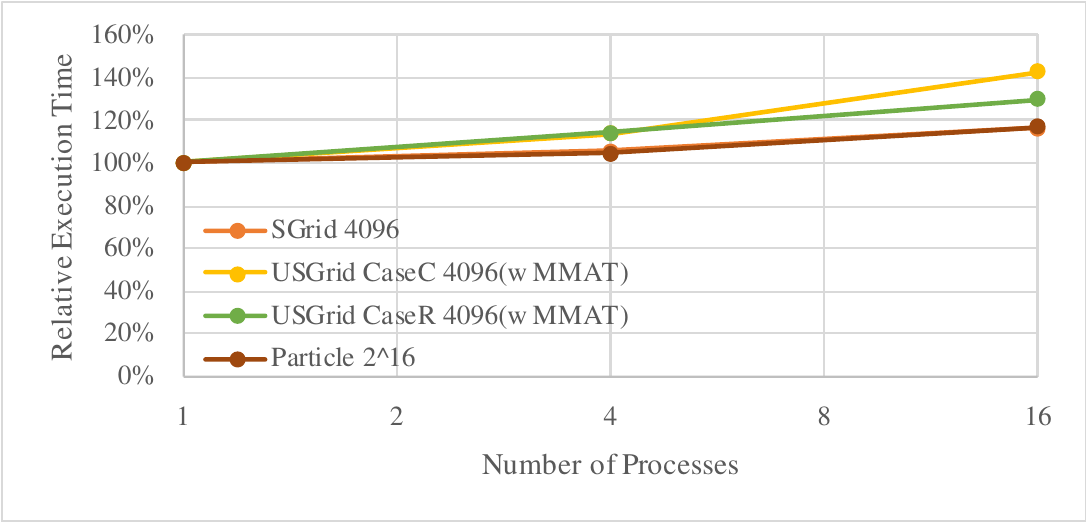}
\caption{Weak scaling performance on OpenMP parallel.}
\label{fig:bench5}
\end{figure}

Fig.~\ref{fig:bench5} indicates a gradual performance degradation is observed in every case.
The performance degradation in CaseC is more significant than that in CaseR.
It might be because, in CaseC, multiple threads simultaneously perform sequential memory accesses to different regions, causing a relatively high number of cache thrashings.

\subsection{MPI+OpenMP}
We measured the performance when 16 tasks were executed using a combination of MPI and OpenMP.
We used MMAT for USGrid samples.

\begin{itemize}
  \item RegionSize (``SGrid'', ``USGrid''): $4096\times4096$ 
  \item Number of Particles (``Particle''): $2^{18}$
  \item MPI Process: One process per node
  \item The number of (MPI Processes$\times$OpenMP Threads): ($1\times16$), ($2\times8$), ($4\times4$), ($8\times2$), ($16\times1$)
\end{itemize}

\begin{figure}
\includegraphics[width=0.95\linewidth]{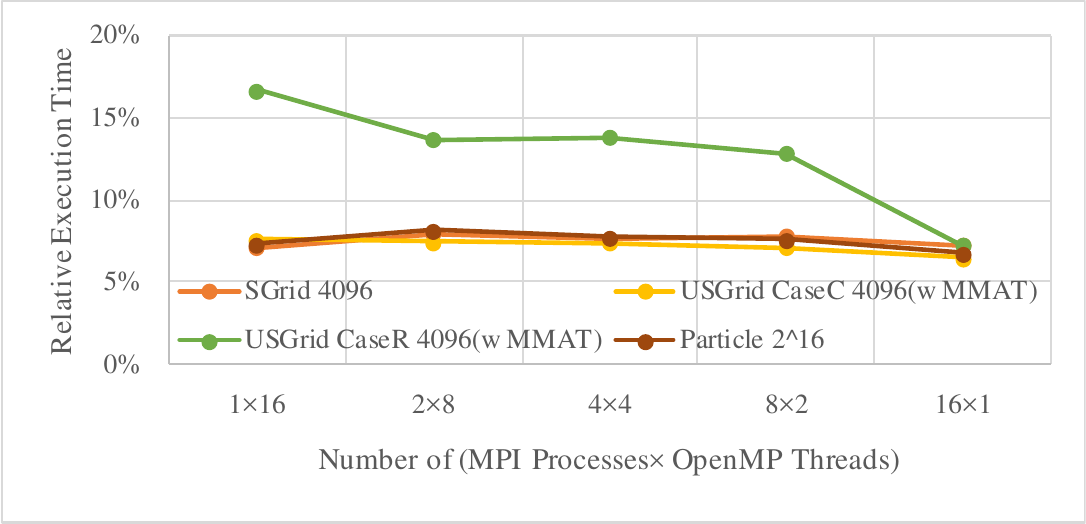}
\caption{Performances on MPI+OpenMP parallel.}
\label{fig:bench6}
\end{figure}

Fig.~\ref{fig:bench6} are normalized the execution time of the execution time on one MPI Process and one OpenMP Thread as 100\%.
The performance of ``USGrid CaseR'' worsened in the case with 16 OpenMP threads, while there was no significant difference in the other cases.
Since ``USGrid CaseR'' does not scale well with OpenMP in terms of parallelism, the results deteriorate as the number of parallel OpenMP threads increases.

\subsection{Memory usage}
\label{sec:memusage}

We measured the amount of memory usage at the execution.
In addition to the cases in section~\ref{sec:overhead}, we compared the size of the case using both MPI and OpenMP (Platform MPI+OMP).
Valgrind\cite{Valgrind28:online} was used to measure memory usage.

\begin{itemize}
  \item RegionSize (``SGrid'', ``USGrid''): $512\times512$ 
  \item Number of Particles (``Particle''): $2^{14}$
  \item \textit{Memory Pool} Size: 300MB
  \item Number of MPI Processes and OpenMP threads: $1$
\end{itemize}

\begin{figure}
  \includegraphics[width=0.95\linewidth]{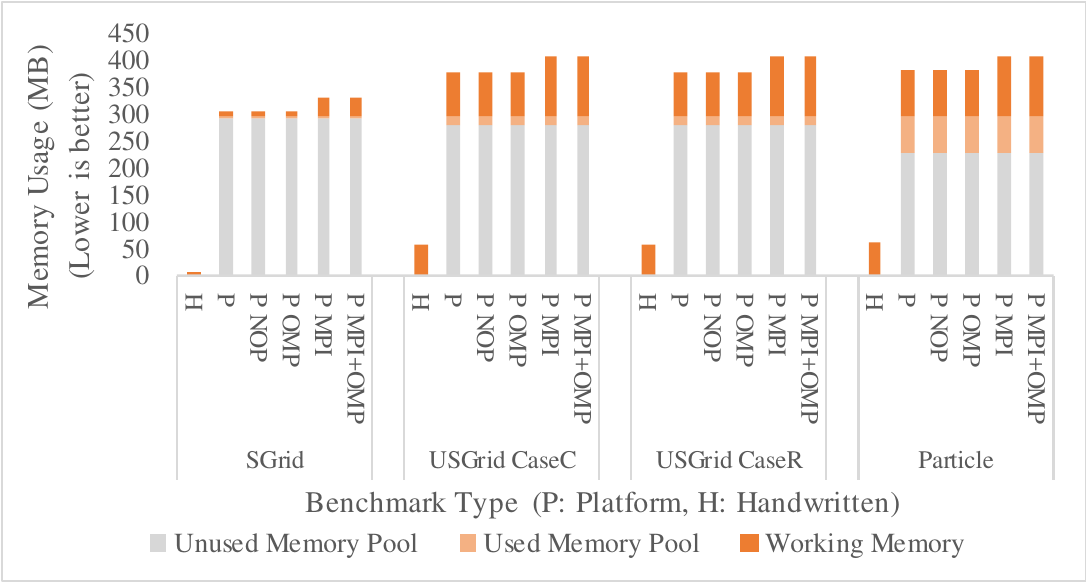}
  \caption{Comparison of total memory usage.}
  \label{fig:bench7}
\end{figure}

Fig.~\ref{fig:bench7} shows the size of the unused memory pool, used memory pool, and other working memory with separate colors.
Even if excluding fixed-size memory pools, the memory usage of the cases with the platform is larger several to dozens of times.
It is due to data of the structure of Env and MMAT.

\subsection{Binary size}

We measured the size of benchmark programs used in section~\ref{sec:memusage}.
The results of USGrids are unified because USGrid CaseC and CaseR are the same binary.

\begin{table}[]
  \centering
  \caption{The size of binaries of benchmarks (KB). (P: Platform, H: Handwritten)}
  \label{tab:size}
\begin{tabular}{ccccccc}
 & H & P & P NOP & P OMP & P MPI & P MPI+OMP \\
 \hline
SGrid & 64 & 199 & 201 & 216 & 242 & 249 \\
\hline
USGrid & 78 & 200 & 204 & 212 & 245 & 250 \\
\hline
Particle & 44 & 191 & 193 & 207 & 234 & 241
\end{tabular}
\end{table}

Table~\ref{tab:size} indicates that the size of the binaries with the platform is three to five times larger but still within the size of the CPU cache.

\subsection{Amount of code}

At last, we counted the number of lines of each benchmark.

\begin{table}
\centering
\caption{The amount of code lines of benchmarks without blank lines and comments. (P: Platform, H: Handwritten)}
\label{tab:code}
\begin{tabular}{ccccccc}
    & \multicolumn{2}{c}{SGrid} & \multicolumn{2}{c}{USGrid} & \multicolumn{2}{c}{Particle} \\
& P & H & P & H & P & H \\
\hline
Platform Part & 3224 & - & 1099 & - & 1099 & - \\
\hline
DSL Part & 466 & - & 360 & - & 597 & - \\
\hline
App Part & 76 & 59 & 139 & 133 & 139 & 176 \\
\end{tabular}
\end{table}

Table~\ref{tab:code} indicates that the amount of code, which end-users should write, is about the same as that of handwritten.
Due to the increase in DSL library notation, the need to call functions provided by the library, and the decrease in the need for initialization and completion cancel each other out.

\section{Conclusion}
By creating a prototype DSL constructing platform, we built several test DSL processing systems and confirmed that they could be parallelized using a combination of the aspect module provided by the platform.
We evaluated the performance and confirmed that although it deteriorated on a single node, it was better than a handwritten serial program on more than two tasks if benchmarks meet the assumptions.
Since most current computers are equipped with processors with two or more cores, the advantages of parallelization can be obtained by simply creating a serial program on the platform, and performance improvements can be obtained.
Therefore, we confirmed that our design is helpful as a DSL constructing platform to improve productivity for DSL developers and end-users.
However, performance degradation can occur when porting the program that uses a considerable amount of memory relative to the amount of memory on the system caused by not efficient memory efficiency of the platform.

As future work, we plan to extend the platform features described below.
\begin{itemize}

\item \textit{Subkernel} modification: In the prototype, the same \textit{subkernel} runs on all tasks on a homogeneous processor and performs these two processes.
However, the \textit{subkernel} and processor are not necessarily homogeneous if they can execute each process.
Then we plan to implement an internal DSL for a \textit{subkernel}, and the platform generates \textit{kernels} for multiple types of processors and executes them heterogeneously, using GPUs, SIMD, and other accelerators.

\item Cache of data access resolution: On current platforms, each memory access involves resolving the address to which the memory is accessed; multiple identical accesses even within a single \textit{subkernel} incur unnecessary address resolution costs. Then, performance will improve by reusing the address resolution.
Also, by reordering the instruction sequence, memory accesses can be made sequential, improving performance.

\end{itemize}

In order to achieve the above two, we are considering using a JIT compilation of the \textit{subkernel} part.

\section*{acknowledgment}
This research was conducted using the Fujitsu PRIMERGY CX400M1/CX2550M5 (Oakbridge-CX) in the Information Technology Center, The University of Tokyo.
The computational resource of Oakbridge-CX was awarded by the ``Computational Science Alliance'' Project, The University of Tokyo.

\bibliographystyle{IEEEtran}
\bibliography{IEEEabrv, mybib.bib}

\end{document}